\documentclass[journal]{IEEEtran}

\usepackage{cite}
\ifCLASSINFOpdf
\else
   \usepackage[dvips]{graphicx}
\fi

\usepackage{url}
\usepackage{graphicx}
\usepackage{amsmath}
\usepackage{hyperref}
\usepackage{booktabs}
\usepackage{array}

\newcolumntype{P}[1]{>{\centering\arraybackslash}p{#1}}
\newcolumntype{M}[1]{>{\centering\arraybackslash}m{#1}}
\usepackage{bm}
\usepackage{amssymb}
\usepackage{amsmath}
\usepackage{color}
\usepackage{multirow}
\usepackage{tablefootnote}

\usepackage{xcolor}

\providecommand{\zd}[1]{\textcolor{red}{{#1}}}

\begin{document}

\title{One-class Learning Towards Synthetic Voice Spoofing Detection}

\author{You Zhang, \IEEEmembership{Student Member, IEEE}, Fei Jiang, and Zhiyao Duan, \IEEEmembership{Member, IEEE}
\thanks{This work was supported by National Science Foundation grant No. 1741472 and funding from Voice Biometrics Group.}
\thanks{You Zhang and Zhiyao Duan are with the Department of Electrical and Computer Engineering, University of Rochester, Rochester, NY 14627 USA (e-mails: \{you.zhang, zhiyao.duan\}@rochester.edu), Fei Jiang is with the Beijing Institute of Technology, Beijing 100811, China, and also with the University of Rochester, Rochester, NY 14627 USA (e-mail: flyjiang92@gmail.com).}
}

\markboth{Journal of \LaTeX\ Class Files, Vol. 14, No. 8, August 2015}
{Shell \MakeLowercase{\textit{et al.}}: Bare Demo of IEEEtran.cls for IEEE Journals}
\maketitle

\begin{abstract}
Human voices can be used to authenticate the identity of the speaker, but the automatic speaker verification (ASV) systems are vulnerable to voice spoofing attacks, such as impersonation, replay, text-to-speech, and voice conversion. 
Recently, researchers developed anti-spoofing techniques to improve the reliability of ASV systems against spoofing attacks. However, most methods encounter difficulties in detecting unknown attacks in practical use, which often have different statistical distributions from known attacks. Especially, the fast development of synthetic voice spoofing algorithms is generating increasingly powerful attacks, putting the ASV systems at risk of unseen attacks.
In this work, we propose an anti-spoofing system to detect unknown synthetic voice spoofing attacks (i.e., text-to-speech or voice conversion) using one-class learning. The key idea is to compact the bona fide speech representation and inject an angular margin to separate the spoofing attacks in the embedding space. Without resorting to any data augmentation methods, our proposed system achieves an equal error rate (EER) of 2.19\% on the evaluation set of ASVspoof 2019 Challenge logical access scenario, outperforming all existing single systems (\textit{i.e.}, those without model ensemble).
% \NZ{The word of ``unseen'' and ``unknown'' seems confusing, shall we use ``unseen'' through all the places?}
\end{abstract}


\IEEEpeerreviewmaketitle

\section{Introduction}
\label{sec:intro}

\IEEEPARstart{S}{peaker} verification plays an essential role in biometric authentication; it uses acoustics features to verify whether a given utterance is from the target person~\cite{delac2004survey}. 
%The given utterances are usually expected to be genuine speech. 
However, ASV systems can be fooled by spoofing attacks, such as impersonation (mimics or twins), replay (pre-recorded audio), text-to-speech (converting text to spoken words), and voice conversion (converting speech from source speaker to target speaker)~\cite{wu2015survey, todisco2019asvspoof}.
% Among the artificial attacks, replay attacks are called \textit{physical access} (PA) attacks, while text-to-speech and voice conversion attacks are called \textit{logical access} (LA) attacks as the input speech are synthetic~\cite{todisco2019asvspoof}. 
% \nz{Referring to how the spoofing attacks are presented, only pre-recorded genuine speech is replayed to form \textit{physical access} (PA) attacks, while the synthetic speech generated by text-to-speech and/or voice conversion algorithms is directly injected into the system to form \textit{logical access} (LA) attacks~\cite{todisco2019asvspoof}.}
Among them, synthetic voice attacks (including text-to-speech (TTS) and voice conversion (VC)) 
% techniques are usually employed to generate synthetic spoofing attacks, where the speech signals are directly injected into the ASV system~\cite{todisco2019asvspoof}.
% \zd{Such speech synthesis technologies} 
are posing increasingly more threats to speaker verification systems due to the fast development of speech synthesis techniques~\cite{kamble2020advances, das2020predictions}.

To improve the spoofing-robustness of speaker verification systems, stand-alone anti-spoofing modules are developed to detect spoofing attacks.
%Anti-spoofing refers to the research field that develop methods to detect whether the input speech is from a real person rather than spoofing attacks. 
The ASVspoof challenge series ~\cite{wu2015challenge2015, Kinnunen2017, todisco2019asvspoof} has been providing datasets and metrics for anti-spoofing speaker verification research.

%The 2015 and logical access part of 2019 challenge care about AI-generated speech detection.
% Similar to speaker verification, anti-spoofing uses some acoustic features to verify the genuineness of the given utterances. 

In this paper, we focus on anti-spoofing synthetic speech attacks, i.e., discriminating bona fide speech from those generated by TTS and VC algorithms.
Traditional methods pay much attention to feature engineering, where good performance has been shown with hand-crafted features such as Cochlear Filter Cepstral Coefficients Instantaneous Frequency (CFCCIF)~\cite{patelwinner2015}, Linear Frequency Cepstral Coefficients (LFCC)~\cite{sahidullah2015comparison}, and Constant-Q Cepstral Coefficients (CQCC)~\cite{CQCC2016}. 
% CQCC is extracted based on CQT, which reflects human perception system ~\cite{CQCC2016}. It shows good generalization ability across datasets~\cite{todisco2017constant}.  
% LFCC and CQCC features were employed in the baseline systems of the ASVspoof 2019 Challenge~\cite{todisco2019asvspoof}.
As for the back-end classifier, Gaussian Mixture Model (GMM) is used in traditional methods~\cite{patelwinner2015, sahidullah2015comparison, CQCC2016, relativephaseinfo2015, sanchezphase2015}. 
% however, GMM is \zd{shown} not robust for clustering high dimensional features~\cite{tian2016}. 
% With the prevalence of deep learning, some basic deep learning models like have shown good performance in spoofing detection. 
Zhang et al.~\cite{deeplearning2017} investigated deep learning models for anti-spoofing and proved that combinations of Convolutional Neural Networks (CNNs) and Recurrent Neural Networks (RNNs) can improve system robustness. Monteiro et al.~\cite{monteiro2020e2e} proposed to adopt the deep residual network (ResNet) with temporal pooling. Chen et al.~\cite{chen2020} proposed a system which employs the ResNet with large margin cosine loss and applied frequency mask augmentation. Gomez-Alanis et al.~\cite{gomez2019light} adapted a light convolutional gated RNN architecture to improve the long-term dependency for spoofing attacks detection. Wu et al.~\cite{wu2020light} proposed a feature genuinization based light CNN system that outperforms other single systems for detection of synthetic attacks. Aravind et al.~\cite{aravind2020audio} explored transfer learning approach with a ResNet architecture. 
% \NZ{Shall we expand this?}
% subband model and fusion and augmentation
% Inspired by subband process in the feature extraction~\cite{tak2020explainability}, researchers recently introduced subband modeling for the back-end model. Chetti et al~\cite{chettri2020} proposed a subband modeling method where they employed several subnetworks to combine artifacts detection from frequency subbands. Tak et al.~\cite{tak2020spoofing} shows non-linear fusion provides good performance. Some other works~\cite{tian2016, STC2019} also showed that model fusion could improve the performance. \NZ{These subband modeling section could be pruned?} 
% Inspired by sub-band processing in the feature extraction,   and score fusion for the backend model, to combine artifacts detection from frequency sub-bands.
To further improve the performance, researchers introduced model fusion based on sub-band modeling~\cite{tak2020spoofing} and different features~\cite{tian2016, STC2019, chettri2019ensemble} at the cost of increased model complexity.
%However, the fusion methods \zd{increased} the model complexity \zd{significantly}.

%% Motivation: why two class not work for anti-spoofing and one-class work
While much progress has been made, existing methods generally suffer from generalization to unseen spoofing attacks in the test stage~\cite{wu2015survey, kamble2020advances}. We argue that this is because most methods formulate the problem as binary classification of bona fide and synthetic speech, which intrinsically assumes the same or similar distributions between training and test data for both classes. %When unseen spoofing attacks comes in the test stage, the systems usually suffer from generalization issue~\cite{wu_survey2015, kamble2020advances}.
While this assumption is reasonable for the bona fide class given a big training set with diverse speakers, it is hardly true for the fake class. Due to the development of speech synthesis techniques, the synthetic spoofing attacks in the training set may never be able to catch up with the expansion of the distribution of spoofing attacks in practice. %Therefore, the binary classification formulation cannot satisfy the need for the generalization ability of anti-spoofing systems.

This distribution mismatch between training and test for the fake class actually makes the problem a good fit for one-class classification. In the one-class classification setup~\cite{khan2009survey}, there is a target class that does not have this distribution mismatch problem, while for the non-target class(es), samples in the training set are either absent or statistically unrepresentative. The key idea of one-class classification methods is to capture the target class distribution and set a tight classification boundary around it, so that all non-target data would be placed outside the boundary.% could be mapped outside the genuine data in the feature space

% Figure~\ref{fig:oneclass} \zd{illustrates the ideas of binary classification and one-class classification.} For binary classification, the decision boundary well separates the training data, but it lacks the ability to detect unseen non-target data \zd{due to its distribution mismatch between training and testing}. \zd{The one-class classification boundary is much tigher around the target class, and successfully detects unseen distributions of non-target data in the test set.}

%To solve one-class problem, the goal is to capture the target data descriptions and set a tight boundary around the target data, so that all the non-target data could be mapped outside the genuine data in the feature space, as shown in Figure~\ref{fig:oneclass} (b).

%% Related one-class work.
% In the field of anti-spoofing, the spoofing attacks keep up-to-date, so the known attacks do not form the exact distribution \fj{of the non-target class}.
% The key issue in one-class classification is to find 1) high-dimensional representation of the data, encouraging the compactness of the target data and descriptiveness between target and non-target data, and 2) a well-chosen metric to indicate the confidence whether the data belong to the target class. \NZ{This sentence may be unrelated?}
% Three categories of one-class methods
% a. Data description method
% b. Generates outlier data around the target set.
% c. Make use of reference data
Recently, the one-class learning idea has been successfully introduced into image forgery detection ~\cite{khalid2020oc, baweja2020anomaly, masi2020two}.
% Perera et al.~\cite{perera2019deep} introduced membership loss and negative filters for novelty detection. They further extend their work to a training scheme~\cite{perera2019learning} with compactness loss and descriptive loss. 
% Khalid et al.~\cite{khalid2020oc} proposed to use variational autoencoder to learn the latent space of natural data and use reconstruction error as scoring. Baweja et al.~\cite{baweja2020anomaly} proposed an adaptive mean estimation strategy to generate pseudo-negative data to address the close proximity between attacks and target data. These methods only took advantage of genuine data, but for voice spoofing detection, it is very likely that most information the model covers are unrelated to genuineness, since speech contains complex information like language, speaker, gender, etc. Masi et al.~\cite{masi2020two} introduced a novel loss that induced the compactness of the natural videos and encourage isolating manipulated ones. 
For voice spoofing detection, Alegre et al.~\cite{alegre2013one} employed a one-class support vector machine (OC-SVM) trained only on bona fide speech to classify local binary patterns of speech cepstrograms, showing the potential of one-class classification approach. It did not make use of any information from spoofing attacks. Villalba et al. \cite{villalba2015spoofing} proposed to fit OC-SVM with DNN extracted speech embeddings of the bona fide class for the ASVspoof 2015 Challenge. Although the method uses OC-SVM to learn a compact boundary for the bona fide class, the embedding space is still learned through binary classification. In other words, the embedding space for drawing the classification boundary may not benefit one-class classification. 
% \NZ{The using of ``features'' may cause ambiguity, since LFCC are also features. How about using ``embedding''?}\ZD{Fixed.}
%However, we argue that the two-stage training leads that the learned bottleneck features are not the best for scoring. The features are still learned from distinguishing known attacks and set boundary with OC-SVM. 
%The separate feature learning process may limit the performance when detecting unknown attacks. 

In this paper, we formulate the speaker verification anti-spoofing problem as one-class feature learning to improve the generalization ability. The target class refers to bona fide speech and the non-target class refers to spoofing attacks. We propose a loss function called one-class softmax (OC-Softmax) to learn a feature space in which the bona fide speech embeddings have a compact boundary while spoofing data are kept away from the bona fide data by a certain margin.
Our proposed method, without resorting to any data augmentation, outperforms all existing single systems (those without ensemble learning) on the ASVspoof 2019 LA dataset, and ranks between the second and third places among all participating systems. 

\section{Method}
\label{sec:method}

Typically, for deep-learning based voice spoofing detection models, the speech features are fed into a neural network to calculate an embedding vector for the input utterance. The objective of training this model is to learn an embedding space in which the bona fide voices and spoofing voices can be well discriminated. The embedding would be further used for scoring the confidence of whether the utterance belongs to bona fide speech or not. To the best of our knowledge, all the previous voice spoofing detection systems learn the speech embedding using a binary classification loss function. As discussed in Section \ref{sec:intro}, this may limit the generalization ability to unknown attacks as spoofing algorithms evolve. In this section, we first briefly introduce and analyze the widely used binary classification loss functions, then propose our one-class learning loss function for voice spoofing detection.

\begin{figure}[]
  \centering
  \centerline{\includegraphics[width=8.5cm]{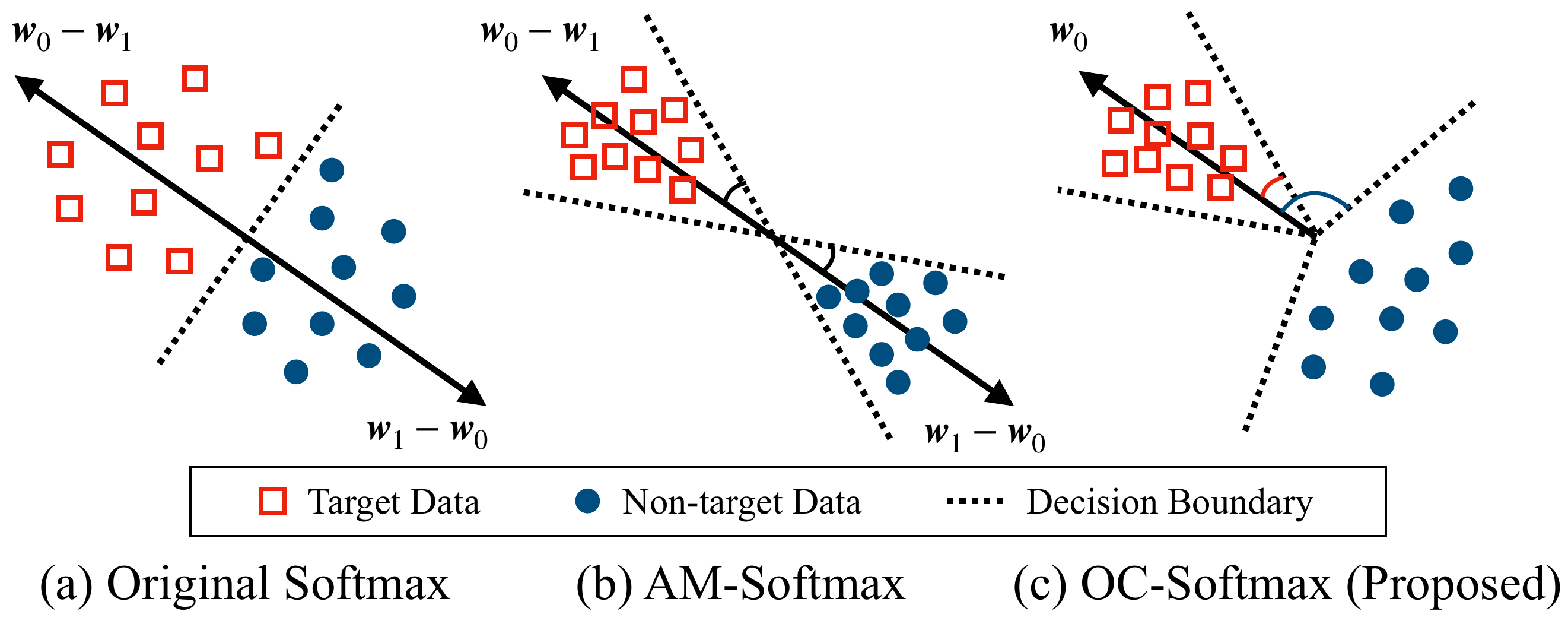}}
% \begin{minipage}[b]{.48\linewidth}
%   \centering
%   \centerline{\includegraphics[height=2.8cm]{Figure/two-class_v2.jpg}}
% %  \vspace{1.5cm}
%   \centerline{(a) Binary classification}\medskip
% \end{minipage}
% \hfill
% \begin{minipage}[b]{0.48\linewidth}
%   \centering
%   \centerline{\includegraphics[height=2.8cm]{Figure/one-class_v2.jpg}}
% %  \vspace{1.5cm}
%   \centerline{(b) One-class classification}\medskip
% \end{minipage}
%
\caption{Illustration of the original Softmax and AM-Softmax for binary classification, and our proposed OC-Softmax for one-class learning. (The embeddings and the weight vectors shown are non-normalized.)}
\label{fig:ocsoftmax}
\end{figure}

\subsection{Preliminary: binary classification loss functions}
\label{ssec:biloss}
% To better understand our proposed loss function, the original softmax and additive margin softmax (AM-Softmax) would be briefly reviewed here. The have good discriminative ability in binary classification, but have limitations in one-class learning.
The original Softmax loss for binary classification can be formulated as
\begin{equation}
\label{eq:softmax}
\begin{aligned}
\mathcal L_{S}&=-\frac{1}{N}\sum_{i=1}^N\log \frac{e^{\bm w_{y_i}^T \bm x_i}}{e^{\bm w_{y_i}^T\bm x_i}+e^{\bm w_{1-y_i}^T\bm x_i}}\\
&=\frac{1}{N}\sum_{i=1}^N\log\big (1+e^{(\bm w_{1-y_i}-\bm w_{y_i})^T\bm x_i}\big),
\end{aligned}
\end{equation}
where $\bm x_i \in \mathbb{R}^{D}$ and $y_i \in \{0, 1\}$ are the embedding vector and label of the $i$-th sample respectively, $\bm w_0, \bm w_1 \in \mathbb{R}^{D}$ are the weight vectors of the two classes, and $N$ is the number of samples in a mini-batch. %The objective of this loss function is to maximize the inner product between $\bm w_{y_i}-\bm w_{1-y_{i}}$ and $\bm x_i$.

% AM-Softmax is originally proposed in~\cite{wang2018additive}. It aims to force the deep neural network to learn the feature embedding that can maximize the inter-class variance and minimize the intra-class variance, by reforming the softmax loss as a cosine loss and injecting a margin in the cosine space.

AM-Softmax~\cite{wang2018additive} improves upon this by introducing an angular margin to make the embedding distributions of both classes more compact, around the weight difference vector's two directions:
\begin{equation}
\label{eq:am-softmax}
\begin{aligned}
\mathcal L_{AMS}&=-\frac{1}{N}\sum_{i=1}^N\log \frac{e^{\alpha(\hat{\bm w}_{y_i}^T \hat{\bm x}_i-m)}}{e^{\alpha(\hat{\bm w}_{y_i}^T\hat{\bm x}_i-m)}+e^{\alpha\hat{\bm w}_{1-y_i}^T\hat{\bm x}_i}}\\
&=\frac{1}{N}\sum_{i=1}^N\log\Big (1+e^{\alpha\big(m-(\hat{\bm w}_{y_i}-\hat{\bm w}_{1-y_i})^T\hat{\bm x}_i\big)}\Big),
\end{aligned}
\end{equation}
where $\alpha$ is a scale factor, $m$ is the margin for cosine similarity, and $\hat{\bm w}$, $\hat{\bm x}$ are normalized $\bm w$ and $\bm x$ respectively.

\subsection{Proposed loss function for one-class learning}
\label{ssec:oneloss}
% \textbf{One-class SVM}:~\cite{scholkopf2000support} 

% \begin{equation}\label{eq:ocsvm}
% \begin{aligned}
% L=\frac{1}{2}\|w\|_{2}^{2} +\frac{1}{\nu} \cdot \frac{1}{P} \sum_{n=1}^{P} \max \left(0, r-\left\langle w, \Phi\left(I_{i}\right)\right\rangle\right)-r\\
% \end{aligned}
% \end{equation}
% \NZ{Do we need this? If so, we also need to compare with SVDD~\cite{tax2004support}}

% \textbf{Isolate Loss}: This loss was originally proposed in ~\cite{masi2020two} for DeepFake detection. Their motivation is to induce the compactness of the embedding space for unmanipulated faces by closing to a center and and encourage isolating manipulated ones. \NZ{Our center is learned, different from their paper, where to mention this? Here or the experimental implementation?}
% \begin{equation}\label{eq:iso}
% \begin{aligned}
% L&=\frac{1}{P} \sum_{i}^{P} \max \left(0, \left\|\Phi\left(I_{i}\right)-c\right\|_{2}-r^{-}\right)\\
% &+\frac{1}{N} \sum_{j}^{N}  \max \left(0, r^{+}-\left\|\Phi\left(I_{j}\right)-c\right\|_{2}\right)
% \end{aligned}
% \end{equation}
% The center $c$, small radius $r^{-}$, and large radius $r^{+}$ are pre-set hyper-parameters. If the genuine data is outside the $r^{-}$, it will be pull back. If the spoofing data is inside the large radius, it wold be push far away. The margin $r^{+}-r^{-}$ ensures the discriminative ability to the model. For the scoring, we measure the distance with the center in the embedding space.

According to the formulae of Softmax and AM-Softmax in Eq. \eqref{eq:softmax} and Eq. \eqref{eq:am-softmax}, for both loss functions, the embedding vectors of the target and non-target class tend to converge around two opposite directions, i.e., $\bm w_0-\bm w_1$ and $\bm w_1-\bm w_0$, respectively. This is shown in Fig. \ref{fig:ocsoftmax} (a), (b). For AM-Softmax, the embeddings of both target and non-target class are imposed with an identical compactness margin $m$. The larger $m$ is, the more compact the embeddings will be. 

In voice spoofing detection, it is reasonable to train a compact embedding space for bona fide speech. However, if we also train a compact embedding space for the spoofing attacks, it may overfit to known attacks. 
To address this issue, 
% we introduce one-class learning to treat genuine speech and spoofing attacks differently. W
we propose to introduce two different margins for better compacting the bona fide speech and isolating the spoofing attacks. The proposed loss function One-class Softmax (OC-Softmax) is denoted as 
% \begin{equation}
% \label{eq:ang_iso}
% \mathcal L=\log \big(1+\sum_{y_i=0}e^{\alpha(m_0-\bm w_0 \bm x_i)}+\sum_{y_i=1}e^{\alpha(\bm w_0\bm x_i-m_1)}\big).
% \end{equation}
% \begin{equation}
% \label{eq:oc-softmax}
% \mathcal L_{OCS}=\frac{1}{N}\sum_{i=1}^N\log \big(1+e^{\alpha(m_{y_i}-\bm w_0\bm x_i)(-1)^{y_i}} \big).
% \end{equation}
\begin{equation}
\label{eq:oc-softmax}
\begin{aligned}
\mathcal L_{OCS}=\frac{1}{N}\sum_{i=1}^N\log \big(1+e^{\alpha(m_{y_i}-\hat{\bm w}_0\hat{\bm x}_i)(-1)^{y_i}} \big).
\end{aligned}
\end{equation}
Note that only one weight vector $\bm w_0$ is used in this loss function. The $\bm w_0$ refers to the optimization direction of the target class embeddings. The $\bm w_0$ and $\bm x$ are also normalized as in AM-Softmax. Two margins ($m_0, m_1 \in[-1, 1], m_0 > m_1$) are introduced here for bona fide speech and spoofing attacks respectively, to bound the angle between $\bm w_0$ and $\bm x_i$, which is denoted by $\theta_i$. When $y_i=0$, $m_0$ is used to force $\theta_i$ to be smaller than $\arccos m_0$, whereas when $y_i=1$, $m_1$ is used to force $\theta_i$ to be larger than $\arccos m_1$. As shown in Fig. \ref{fig:ocsoftmax} (c), a small $\arccos m_0$ can make the target class concentrate around the weight vector $\bm w_0$, whereas a relatively large $\arccos m_1$ can push the non-target data to be apart from $\bm w_0$.

% Another recently proposed isolate loss is recently proposed in \cite{masi2020two} for DeepFake detection. Our proposed angular isolate loss is different in the sense that we use cosine similarity as the metric to measure the distance to the center. 
% Compared with Isolate loss, our angular loss has less parameters. The angle between feature vectors are robust compared to Euclidean distance. More soft

% \begin{equation}\label{eq:ocnn_loss}
% \begin{aligned}
% L=\min _{w, V, r} \frac{1}{2}\|w\|_{2}^{2}&+\frac{1}{2}\|V\|_{F}^{2} \\ &+\frac{1}{\nu} \cdot \frac{1}{N} \sum_{n=1}^{N} \max \left(0, r-\left\langle w, g\left(V \mathbf{X}_{n:}\right)\right\rangle\right)-r\\
% \end{aligned}
% \end{equation}

\section{Experiments}
\label{sec:experiments}

\subsection{Dataset}
\label{ssec:dataset}
The ASVspoof 2019 challenge provides a standard database \cite{dataset2020} for anti-spoofing. The LA subset of the provided dataset includes bona fide speech and different kinds of TTS and VC spoofing attacks. Training and Development sets share the same 6 attacks (A01-A06), consisting of 4 TTS and 2 VC algorithms. %In the training and validation sets, there are 6 attacks (A01-A06) consisting of 4 TTS and 2 VC algorithms. 
In the test set, there are 11 unknown attacks (A07-A15, A17, A18) including combinations of different TTS and VC attacks. The test set also includes two attacks (A16, A19) which use the same algorithms as two of the attacks (A04, A06) in the training set but trained with different data.
% \NZ{The dataset and challenge paper call these two known attacks, but I prefer to call them similar attacks but still unknown. Would that be a issue?} \ZD{Let's follow the challenges definition and explain it as what you did.}
Details can be found in Table \ref{tab:dataset}. The TTS and VC algorithms in the dataset are mostly based on neural networks.

\subsection{Evaluation metrics}
\label{ssec:evaluate}
To evaluate the performance of the anti-spoofing system, we take note of the output score of the anti-spoofing system. The output of the anti-spoofing system is called a countermeasure (CM) score, and it indicates the similarity of the given utterance with bona fide speech. 
% \subsubsection{Equal Error Rate (EER)}\label{subsubsec:eer}
Equal Error Rate (EER) is calculated by setting a threshold on the countermeasure decision score such that the false alarm rate is equal to the miss rate. The lower the EER is, the better the anti-spoofing system is at detecting spoofing attacks.
% \begin{equation}\label{eq:eer}
% \begin{aligned} P_{f a}(\theta)&= \frac{\#\{\text { spoof trials with score }>\theta\}}{\#\{\text { total spoof trials }\}} \\ P_{\text {miss}}(\theta) &=\frac{\#\{\text { human trials with score } \leq \theta\}}{\#\{\text { total human trials }\}} \end{aligned}
% \end{equation}
% \subsubsection{tandem detection cost function (t-DCF)}\label{subsubsec:tdcf}
The tandem detection cost function (t-DCF)~\cite{kinnunen2018} is a new evaluation metric adopted in the ASVspoof 2019 challenge. While EER only evaluates the performance of the anti-spoofing system, the t-DCF assesses the influence of anti-spoofing systems on the reliability of an ASV system. The ASV system is fixed to compare different anti-spoofing systems. The lower the t-DCF is, the better reliability of ASV is achieved.

\subsection{Training details}
\label{ssec:details}
% \subsubsection{Feature Extraction and Model Architecture}
% \label{sssec:feat}
We extract 60-dimensional LFCCs from the utterances with the MATLAB implementation provided by the ASVspoof 2019 Challenge organizers\footnote{\url{https://www.asvspoof.org}}. The frame size is 20ms and the hop size is 10ms. To form batches, we set 750 time frames as the fixed length and use repeat padding for short trials, and we randomly choose a consecutive piece of frames and discard the rest for long trials. 

\begin{table}[]
\caption{Summary of the ASVspoof 2019 LA dataset.}
\begin{tabular}{l|c|cc}
            & Bona fide   & \multicolumn{2}{c}{Spoofed}                         \\ \cline{2-4} 
            & \# utterance & \multicolumn{1}{c|}{\# utterance} & attacks \\ \hline
Training    & 2,580        & \multicolumn{1}{c|}{22,800}       & A01 - A06        \\
Development & 2,548        & \multicolumn{1}{c|}{22,296}       & A01 - A06        \\
Evaluation  & 7,355        & \multicolumn{1}{c|}{63,882}       & A07 - A19       
\end{tabular}
\label{tab:dataset}
\end{table}

% \subsubsection{Model Architecture}
% \label{sssec:arch}
We adopt the network architecture adapted from \cite{monteiro2020e2e}. The architecture is based on deep residual network ResNet-18~\cite{he2016deep}, where the global average pooling layer is replaced by attentive temporal pooling. The architecture takes the extracted LFCC features as input and outputs the confidence score to indicate the classification result. We use the intermediate output before the last fully connected layer as the embedding for the speech utterances, where 256 is the embedding dimension.
% The first convolution layer has kernel size 9$\times$3, stride 3$\times$1, and padding 1.The filter size of 4 Res blocks are 3$\times$3. The strides for the blocks are 2,2,2,2 respectively. The number of filters are 64, 128, 256, For 512. \NZ{Need modify}
For the hyper-parameters in the loss functions, we set $\alpha = 20$ and $m = 0.9$ for AM-Softmax; we set $\alpha = 20$, $m_0 = 0.9$ and $m_1 = 0.2$ for the proposed OC-Softmax.

% \subsubsection{Training Details}
% \label{sssec:train}
We implement our model with PyTorch\footnote{Code is available at \url{https://github.com/yzyouzhang/AIR-ASVspoof}}.
We use Adam optimizer with the $\beta_1$ parameter set to 0.9 and the $\beta_2$ parameter set to 0.999 to update the weights in the ResNet model. We use Stochastic Gradient Descent (SGD) optimizer for the parameters in the loss functions. The batch size is set to 64. The learning rate is initially set to 0.0003 with 50\% decay for every 10 epochs. We trained the network for 100 epochs on a single NVIDIA GTX 1080 Ti GPU. Then we select the model with the lowest validation EER for evaluation. The output CM score of our system is the cosine similarity between the speech embedding and the weight vector $\bm w_0$ in OC-Softmax.

\subsection{Results}
\label{ssec:results}
\subsubsection{Evaluation of proposed loss function}
\label{sssec:eval}
To demonstrate the effectiveness of the one-class learning method, we compared our proposed OC-Softmax with the conventional binary classification loss functions, under the setting of the same input features and models. The performance of the system trained with different loss functions are compared in Table~\ref{tab:compare_loss} on both the development and evaluation sets of ASVspoof 2019 LA scenario. We also compare the performance on individual unseen attacks in the evaluation set in Table~\ref{tab:individual}.

%%%%%%%%%%%%%%%%%%%%%%%%%%%%%%%%%%%%%%%%%%%%%%%

\begin{table}[!t]
\caption{Results on the development and evaluation sets of the ASVspoof 2019 LA scenario using loss functions in Section~\ref{sec:method}.
}
% increase table row spacing, adjust to taste
\renewcommand{\arraystretch}{1.0}
\centering
% \begin{tabular}{M{55pt} M{25pt} M{30pt} M{25pt} M{35pt}}
% \toprule
\begin{tabular}{c c c c c}
\hline\hline
\multirow{2}{*}{\textbf{Loss}} & \multicolumn{2}{c}{\textbf{Dev Set}}& \multicolumn{2}{c}{\textbf{Eval Set}}\\ \cline{2-5}
& EER (\%) & min t-DCF  & EER (\%) & min t-DCF\\
% \textbf{Loss function} & EER (\%) & t-DCF \\
\hline
Softmax & 0.35 & 0.010 & 4.69 & 0.125\\

AM-Softmax & 0.43 & 0.013 & 3.26 & 0.082\\

\textbf{OC-Softmax} & 0.20 & 0.006 & \textbf{2.19} & \textbf{0.059}\\
% \bottomrule
\hline\hline
\end{tabular}
\label{tab:compare_loss}
\end{table}

The three losses perform similarly on the development set, showing that they have good discrimination ability to detect known attacks. For unknown attacks in the evaluation set, our one-class learning with the proposed loss surpasses the binary classification losses (Softmax and AM-Softmax). In particular, the relative improvement on EER is up to 33\%. As for individual attacks, A17 is the most difficult attack in the evaluation set, and our proposed system shows a significant improvement. In addition, the Softmax and AM-Softmax systems show some overfitting as they show close-to-zero results on certain attacks (e.g., A08, A09) while much worse results on other attacks (e.g., A14, A18).
% This suggests that for spoofing attacks similar to seen attacks, they perform well but not for others. 
In contrast, our system with OC-Softmax achieves universally good performance over all but A17 attacks.

%%%%%%%%%%%%%%%%%%%%%%%%%%%%%%%%%%%%%%%%%%%%%

\begin{table}[!t]
\caption{EER (\%) performance comparison of loss functions on individual attacks of the evaluation set \zd{(all unseen from development)} of the ASVspoof 2019 LA scenario.}
% increase table row spacing, adjust to taste
\renewcommand{\arraystretch}{1.0}
\centering
\begin{tabular}{M{50pt} | M{50pt} M{50pt} M{50pt}}
% \toprule
% \begin{tabular}{c c c}
\hline\hline
\textbf{Attacks} & Softmax & AM-Softmax & OC-Softmax \\
\hline
A07	 & 0.37	 & 0.22	 & 0.12   \\
A08	 & 0.01	 & 0.06	 & 0.18   \\
A09	 & 0.02	 & 0.02	 & 0.12  \\
A10	 & 1.18	 & 0.63	 & 1.14   \\
A11	 & 0.37	 & 0.02	 & 0.12   \\
A12	 & 0.43	 & 0.53	 & 0.47   \\
A13	 & 0.69	 & 0.27	 & 0.22   \\
A14	 & 4.98	 & 0.51	 & 0.69   \\
A15	 & 5.43	 & 0.69	 & 1.40   \\
A16	 & 0.22	 & 0.51	 & 0.33   \\
A17	 & 23.48  & 13.45  & 9.22   \\
A18	 & 0.20	 & 4.27	 & 0.90   \\
A19	 & 1.34	 & 0.86  & 0.90   \\

% \bottomrule
\hline\hline
\end{tabular}
\label{tab:individual}
\end{table}

\begin{figure}[!t]

\centering
\centerline{\includegraphics[width=8.0cm]{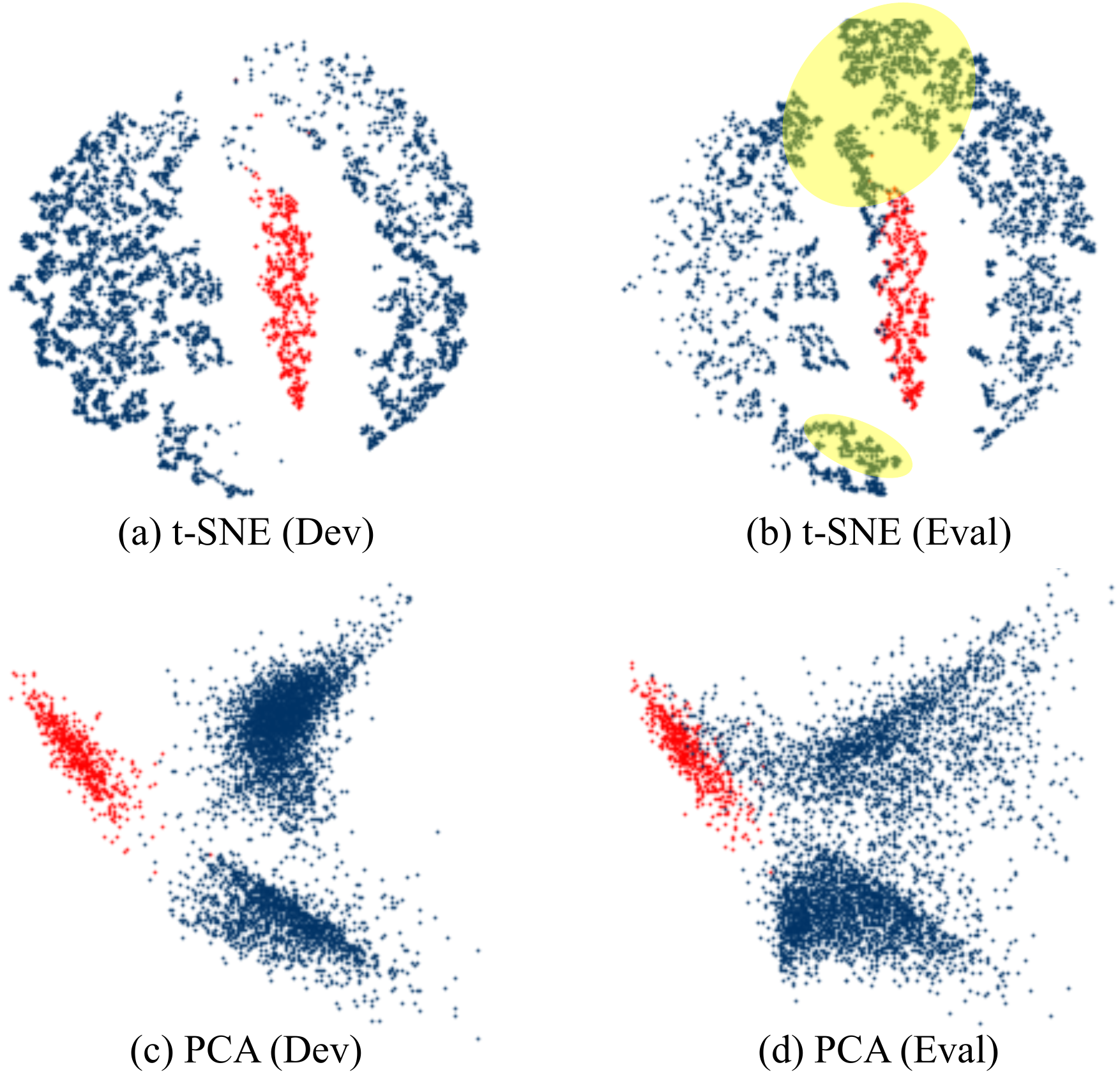}}
\caption{Feature embedding visualization of our proposed loss. Red: Bona fide Speech; Blue: Spoofing Attacks. Highlighted part are the unknown attacks with different distributions.
% The top left shows the t-SNE on the dev set. The top right shows the t-SNE on the eval set. The bottom left shows the PCA on the dev set. The bottom right shows the PCA on the eval set.
% \NZ{How abut putting something like ``t-SNE(dev)'' as subcaptions for the four subfigures?}
}
\label{fig:res}
\end{figure}

Moreover, the dimension reduced embedding visualization is shown in Figure~\ref{fig:res}. The same t-distributed Stochastic Neighbor Embedding (t-SNE) and Principle Component Analysis (PCA) projections are applied to development and evaluation datasets of ASVspoof 2019 LA scenario. In other words, the visualizations of the two sets use the same coordinating systems.
% In the test set visualization, the coordinating system is the same as validation. 
The t-SNE subfigures show that the bona fide speech has the same distribution in both sets while unknown attacks in the evaluation set show different distributions from the known attacks in the development set. This suggests that the bona fide class is well characterized by the instances in the training data, but the spoofing attacks in the training data cannot form a statistical representation of the unknown attacks, hereby verifies our problem formulation in Section~\ref{sec:intro}. Nevertheless, the unknown attacks that appear in the top and bottom clusters (highlighted) on the top right subfigure are successfully separated from the bona fide speech cluster, 
% \NZ{Jianyu suggested to highlight the left cluster in the figure. Shall we put a circle around it?} \ZD{Yes, would be better. Also, is it possible to plot the two figures into the same figure, using different colors/dots, or maybe that's too cluttered.} 
showing a good generalization ability of our system. 

We further verify the one-class idea with PCA visualization of the learned embedding. In the bottom left subfigure, the embeddings of the bona fide speech are compact, and an angular margin is injected between bona fide and spoofing attacks, thanks to the linearity of PCA, verifying our assumption in Figure~\ref{fig:ocsoftmax} (c). The bottom right figure shows that when encountering unknown spoofing attacks, the angle is still maintained, and the embeddings for the unknown attacks are still mapped to the angularly separate space. This shows the effectiveness of our proposed OC-Softmax loss.

%%%%%%%%%%%%%%%%%%%%%%%%%%%%%%%%%%%%%%%%%%%%%%%

\subsubsection{Comparison with other systems}
\label{sssec:compare}
We also compared our system with other existing single systems (no model fusion) without data augmentation in Table~\ref{tab:compare_systems}. For all methods with a reference, we obtained their results from their papers. Some of them participated the ASVspoof 2019 LA challenge and reported better results with model ensembles and data augmentation, but we only compare with their single system version without data augmentation. It is noted that ``ResNet + OC-SVM'' was adapted from~\cite{villalba2015spoofing}, which is the only existing one-class classification method; We replaced their DNN with our ResNet and run the experiments on the ASVspoof 2019 LA dataset for a fair comparison. It can be seen that our proposed system significantly outperforms all of the other systems. In fact, on the leader board of the ASVspoof 2019 Challenge for LA scenario~\cite{todisco2019asvspoof}, our system would rank between the second (EER 1.86\%)~\cite{STC2019} and the third (EER 2.64\%)~\cite{chettri2019ensemble} among all the systems, even though the top three methods all used model fusion. 

%%%%%%%%%%%%%%%%%%%%%%%%%%%%%%%%%%%%%%%%%%%%%

\begin{table}[!t]
\caption{Performance comparison with existing single systems on the evaluation set of the ASVspoof 2019 LA scenario.}
% increase table row spacing, adjust to taste
\renewcommand{\arraystretch}{1.0}
\centering
\begin{tabular}{M{100pt} M{50pt} M{50pt}}
% \toprule
% \begin{tabular}{c c c}
\hline\hline
\textbf{System} & EER (\%) & min t-DCF \\
\hline
CQCC + GMM \cite{todisco2019asvspoof} & 9.57 & 0.237\\
LFCC + GMM \cite{todisco2019asvspoof} & 8.09 & 0.212\\
% \hline
Chettri et al.~\cite{chettri2019ensemble} & 7.66 & 0.179\\
Monterio et al.~\cite{monteiro2020e2e} & 6.38 & 0.142\\
Gomez-Alanis et al.~\cite{gomez2019light} & 6.28 & - \\
Aravind et al.~\cite{aravind2020audio} & 5.32 & 0.151\\
Lavrentyeva et al.~\cite{STC2019} & 4.53 & 0.103\\
ResNet + OC-SVM & 4.44 & 0.115\\
Wu et al.~\cite{wu2020light} & 4.07 & 0.102 \\
Tak et al.~\cite{tak2020spoofing} & 3.50 & 0.090\\
Chen et al.~\cite{chen2020} & 3.49 & 0.092\\

% system4 & - & -\\
% system5 & - & -\\
\textbf{Proposed} & \textbf{2.19} & \textbf{0.059}\\
% \bottomrule
\hline\hline
\end{tabular}
\label{tab:compare_systems}
\end{table}

% \subsection{Detecting Difficult Attacks}
% \label{ssec:difficult}

% \subsection{Ablation Study}
% \label{ssec:ablation}

\section{Conclusion}
\label{sec:conclusion}
In this work, we proposed a voice spoofing detection system based on one-class learning to enhance the robustness of the model against unknown spoofing attacks. 
The proposed system aims to learn a speech embedding space in which bona fide speech has a compact distribution while spoofing attacks reside outside by an angular margin.
% We propose a novel loss function OC-Softmax to compact the genuine speech embeddings and segregate the spoofing attacks.
Experiments showed that the proposed loss outperforms the original Softmax and AM-Softmax that formulate anti-spoofing as a conventional binary classification problem.
The proposed system also outperforms all existing single systems (no model fusion) without data augmentation of the ASVspoof 2019 Challenge LA scenario, and ranks between the second and the third among all participating systems.
% Future work
For future work, we would like to extend our method to detecting other multimedia forgeries.

\section*{Acknowledgment}
%The authors would like to thank the anonymous reviewers for reviewing and giving constructive suggestions to this work. 
We would like to thank Ge Zhu for valuable discussions on speaker verification.

\vfill\pagebreak

\bibliographystyle{IEEEtran}
\bibliography{refs}

\end{document}